*Anisotropie, symétrie, ultrasons /Anisotropy, symmetry, ultrasounds*

**Une nouvelle analyse des symétries d'un matériau élastique anisotrope. Exemple d'utilisation à partir de mesures ultrasonore.**

Marc FRANÇOIS, Yves BERTHAUD et Giuseppe GEYMONAT
*Laboratoire de Mécanique et Technologie,
E.N.S. de Cachan/C.N.R.S./Université Paris-VI
61 Avenue du Président Wilson - 94235 Cachan Cedex (France)*

**Résumé** - Nous traitons dans cette note des tenseurs de rigidité $\mathbb{C}$ mesurés sur des matériaux anisotropes élastiques linéaires dont la symétrie est inconnue (biomécanique, géologie...). Leurs éventuelles symétries (exactes ou non) sont révélées qualitativement par une figure de pôle. La fonction intrinsèque $\mathcal{F}(\mathbb{C},\mathcal{B})$ permet de déterminer le tenseur $\mathbb{C}_{Sym}$, invariant pour le groupe de symétrie $Sym$, le plus proche de $\mathbb{C}$ et sa base naturelle $\mathcal{B}_{Sym}$. Pour chaque niveau de symétrie on peut choisir celle pour laquelle la différence relative entre $\mathbb{C}$ et $\mathbb{C}_{Sym}$ est minimale. Le tenseur $\mathbb{C}$ est ici obtenu par une méthode ultrasonore par contact direct. L'analyse est menée sur un échantillon de bois de chêne.

**A new analysis of the symmetries of an anisotropic elastic material. Exemple from ultrasonic measurements.**

**Abstract** - When symmetries are unknown, experimental techniques give the stiffness tensor $\mathbb{C}$ only in the specimen basis $\mathcal{B}_o$ which is not related to the possible symmetry axis. At first we suggest to use a pole figure to reveal the symmetries of the tensor $\mathbb{C}$. The intrinsic function $\mathcal{F}(\mathbb{C},\mathcal{B})$ allows us to determine the tensor $\mathbb{C}_{Sym}$, invariant by the symmetry group $Sym$, closest to $\mathbb{C}$ and it's natural base $\mathcal{B}_{Sym}$. For each symmetry level we can choose the one which minimizes the discrepancy between $\mathbb{C}$ and $\mathbb{C}_{Sym}$. An ultrasonic experiment is explained and the full analysis is applied to an oak specimen.

*Abridged English version* - It is easier to determine the stiffness tensor $\mathbb{C}$ of an anisotropic material when symmetries are known *a priori*. Otherwise, no simplification is available and the stiffness tensor $\mathbb{C}$ that is measured generally has no symmetry because of experimental errors; it is expressed in the base $\mathcal{B}_o = \{\mathbf{x}^1, \mathbf{x}^2, \mathbf{x}^3\}$ of the specimen which has no relation with possible material symmetry directions.

**Analysis of symmetry.** To have a qualitative idea of the "quasi" symmetries of such a tensor, we define a map based on a crystallographic pole figure. Every normal $\mathbf{r}$ of the half sphere is valued by the relative discrepancy $\mathcal{D}(\mathbf{r})$ (in natural norm) between $\mathbb{C}$ and its symmetric $\mathcal{S}[\mathbf{r}^\perp](\mathbb{C})$ (Auld, 1973). The map (fig. 1) obtained for the stiffness tensor of the measured oak (table 1) shows transverse isotropy (dark band) at first sight and orthotropy when observed much closer (the two spindles).

**Intrinsic determination of the nearest symmetric stiffness tensor**. Let us consider a symmetry group $Sym$ and let $\mathcal{B} = \{\mathbf{r}(1),\mathbf{r}(2),\mathbf{z}\}$ be an associed normed basis : in the case of p-order symmetries $\mathbf{r}(1)$ and $\mathbf{r}(2) = \mathcal{R}[\mathbf{z},2\pi/p](\mathbf{r}(1))$ represent the normals to the first and second plane of symmetry. We then define $\mathcal{F}(\mathbb{C},\mathcal{B})$ as the average of $\mathbb{C}$ on the orbit of *Sym* and corresponding to the base $\mathcal{B}$ (equ. 1). We determine the base $\mathcal{B}_{Sym}$ for which the discrepancy between $\mathbb{C}$ and $\mathcal{F}(\mathbb{C},\mathcal{B})$ is minimum by a simplex method. The obtained tensor $\mathbb{C}_{Sym}$ is then the closest to $\mathbb{C}$ symmetric tensor. As symmetry groups for four-rank tensors are ordered (Cowin, 1987; Yong-Zongh, 1991) we can, for each level of symmetry, determine the best symmetry for $\mathbb{C}$. Choice of the level remains arbitrary. For our specimen, the discrepancy between $\mathbb{C}$ and $\mathbb{C}_{Sym}$ for every symmetry is given in table 3; the superscript corresponds to the symmetry level according to Cowin. So for the second level, orthotropy appears to be the best choice ($\mathbb{C}_{Sym}$ is given table 2) and transverse isotropy the best for the third level.

**Ultrasonic determination of the stiffness tensor.** The 1 MHz transducers are both identical and parallel, coupled by a rod (fig. 2) which allows us to set the polarization angle $\theta_i$. The brief vibration (longitudinal or transversal) along **m** is split in (at least) three possible directions of vibration $\mathbf{u}^i$ by the material and we measure the flying times (so the speeds $V_i$) for the imposed direction of propagation **n**. When sending shear waves (i=2,3), the maximum intensity is reached when $\mathbf{u}^i$ is nearest to the direction $\mathbf{m}^i$ (fig. 3). Under ideal conditions (*complete measure*), we can get the three waves' speeds $V_i$ and the two polarization angles $\theta_2$, $\theta_3$ for quasi-transverse waves (table 4). The remaining quasi longitudinal $\mathbf{u}^1$ is assumed to be orthognal to $\mathbf{u}^2$ and $\mathbf{u}^3$.

The Christoffel's tensor $\mathbf{\Gamma}$ (Love, 1927), can be written from $\mathbb{C}$ in two ways (*equ. 2, 3*). We define the discrepancy tensor **K** (*equ. 4*) (measured values are underlined in the equation) as the difference between these expressions. In order to use all the information available we build here the scalar minimization function j from the norm of **K** (*equ. 5*). The structure of J, sum of every j, allows to separate the variables into two sets: the 81 (21 independents) components of $\mathbb{C}$ and the 2E unknown deflexion angles $\varphi_i$ (E: number of complete measures, E ≥ 6).

The nullity of the derivative of J with respect to the 81 components of $\mathbb{C}$ has a simple linear form allowing a matricial resolution while the symmetries of $\mathbb{C}$ are obtained by 60 Lagrangian multipliers. We minimize each j with respect to the two deflexion angles $\varphi_i$ with a B.F.G.S. method as the gradient is available.

For *incomplete measures*, we define the components $L_{ij}$, as the projection of **K** on $\mathbf{u}^i \otimes \mathbf{u}^j$ (*equ. 6*), corresponding to the measured directions $\underline{\mathbf{m}}^i$ and $\underline{\mathbf{m}}^j$. If the non measurable signal is quasi-transversal, we impose the orthogonality of the $\underline{\mathbf{u}}^i$ in order to define the missing polarization angle $\theta_3$ (of the two solutions, we choose the best one). The expression of h from the $L_{ij}$ (*equ. 7*) corresponds to the previous construction h=1/2 **L** : **L** if the base $\underline{\mathbf{u}}^i \otimes \underline{\mathbf{u}}^j$ is orthogonal.

Minimization respect to the two gradients is done as before. The resolution condition becomes 4E+F≥21, F being the number of incomplete measurements. The result provides a stiffness tensor in the specimen's axes, as for the oak in table 1.

1. INTRODUCTION. - La symétrie matérielle est inconnue. Le tenseur d'élasticité $\mathbb{C}$ obtenu par une expérimentation ne possède en général aucune symétrie en raison des imprécisions de mesure. De plus, il est exprimé dans la base $\mathcal{B}_0 = \{\mathbf{x}^1, \mathbf{x}^2, \mathbf{x}^3\}$ de l'éprouvette qui n'a aucun rapport avec les éventuelles directions privilégiées du matériau.

Pour obtenir une première information qualitative sur les "quasi" symétries de ce tenseur, nous proposons une figure de pôles issue des représentations cristallographiques. Pour obtenir des informations plus quantitatives, nous construisons pour un groupe de symétrie $\mathit{Sym}$ donné et une base associée $\mathcal{B}$, le tenseur $\mathcal{F}(\mathbb{C},\mathcal{B})$ invariant par $\mathit{Sym}$. On recherche ensuite la base $\mathcal{B}_{\mathit{Sym}}$ pour laquelle l'écart relatif entre $\mathbb{C}$ et $\mathcal{F}(\mathbb{C},\mathcal{B})$ est minimal ; cet écart indique l'erreur commise en considérant que $\mathbb{C}$ possède cette symétrie. Nous pouvons ainsi obtenir, pour un niveau de symétrie retenu (fonction du problème traité) le tenseur optimal $\mathbb{C}_{\mathit{Sym}}$ et sa base associée $\mathcal{B}_{\mathit{Sym}}$.

Les mesures permettant de déterminer $\mathbb{C}$ peuvent être mécaniques (Hayes, 1969). La méthode ultrasonore par immersion (Hosten, 1983) est habituellement utilisée lorsque la symétrie matérielle est connue, sinon, elle apporte deux inconnues angulaires par polarisation. La méthode par contact direct, bien que moins précise, fixe la direction de propagation et est utilisable aussi quand aucune symétrie matérielle n'est connue *a priori* (Arts, 1993 ; Klima, 1973 ; Neighbours, 1967). Les améliorations que nous lui apportons ont été testées sur un échantillon de bois (chêne) choisi loin du centre de l'arbre pour assurer l'homogénéité. Le matériau est supposé, dans toute l'étude, anisotrope, élastique linéaire et homogène.

2. FIGURE DE POLES DU TENSEUR DE RIGIDITE - Il s'agit d'une projection stéréographique dans laquelle à chaque direction $\mathbf{r}$ du demi espace $\mathbf{x}^3 > 0$ on associe la décorrélation $\mathcal{D}(\mathbf{r})$ qui représente l'écart relatif (norme euclidienne naturelle) entre $\mathbb{C}$ et son symétrique $\mathcal{S}[\mathbf{r}^\perp](\mathbb{C})$ par rapport au plan $\mathbf{r}^\perp$. Le symétrique de $\mathbb{C}$ peut être calculé rapidement en utilisant les matrices de Bond (Auld, 1973). Nous montrons (fig. 1) la figure de pôles obtenue pour notre tenseur $\mathbb{C}$ expérimental (Tableau 1). A première vue on remarque surtout la bande sombre horizontale qui indique que les plans de normale contenue dans $[\mathbf{x}^1,\mathbf{x}^3]$ sont de bons plans de symétrie, et les deux demi taches donnant le plan de normale $\mathbf{x}^2$ comme plan de symétrie. Ceci correspond à un cas d'isotropie transverse. Un examen plus approfondi fait apparaître deux fuseaux plus denses au niveau des bissectrices de $\mathbf{x}^1$ et $\mathbf{x}^3$. Ceci indique également la tendance orthotrope de ce matériau. A ce stade nous ne pouvons pas choisir la meilleure des symétries apparentes.

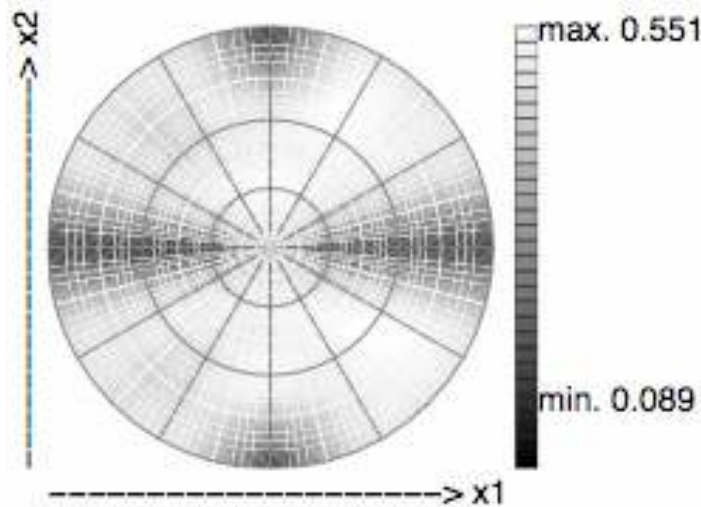

Fig. 1. Figure de pôle de la décorrélation $\mathcal{D}(\mathbf{r})$
*Fig. 1. Pole figure of the discrepancy $\mathcal{D}(\mathbf{r})$.*

$$\mathbb{C} \begin{vmatrix} 3,30 & 0,20 & 1,06 & -0,55 & -0,89 & 0,1 \\ 0,20 & 13,3 & 0,38 & 0,26 & -0,28 & -0,10 \\ 1,06 & 0,38 & 3,04 & 0,07 & -0,05 & -0,43 \\ -0,55 & 0,26 & 0,07 & 0,40 & 0,12 & 0,05 \\ -0,89 & -0,28 & -0,05 & 0,12 & 1,13 & 0,01 \\ 0,1 & -0,10 & -0,43 & 0,05 & 0,01 & 0,76 \end{vmatrix}$$

Tableau 1. Tenseur de rigidité $\mathbb{C}$ brut du chêne (GPa) dans le repère de l'éprouvette.
*Table 1. Raw stifness tensor $\mathbb{C}$ for the oak (GPa) in specimen's axis.*

3. UNE STRATEGIE DE DETERMINATION INTRINSEQUE DU TENSEUR DE RIGIDITE SYMETRIQUE LE PLUS PROCHE - Fixons un groupe de symétrie $Sym$ et soit $\mathcal{B} = \{\mathbf{r}(1),\mathbf{r}(2),\mathbf{z}\}$ une base normée associée : dans le cas des symétries d'ordre p, $\mathbf{z}$ est pris comme axe, le premier plan de symétrie est choisi de normale $\mathbf{r}(1)$ et le second de normale $\mathbf{r}(2) = \mathcal{R}[\mathbf{z},2\pi/p](\mathbf{r}(1))$ ([1]). On peut alors définir $\mathcal{F}(\mathbb{C},\mathcal{B})$ la moyenne de $\mathbb{C}$ sur l'orbite de $Sym$ et correspondante à la base $\mathcal{B}$.

$$(1) \qquad \mathcal{F}(\mathbb{C},\mathcal{B}) = \frac{1}{2p} \sum_{q=0}^{2p-1} \mathcal{S}[\mathbf{r}(2-\text{mod}_2(q))^\perp]^q (\mathbb{C})$$

dans laquelle "$\text{mod}_2$" représente l'opérateur modulo 2, et la puissance q sur l'opérateur de symétrie $\mathcal{S}$ est entendue en terme de composition. Par convention $\mathcal{S}[\mathbf{r}(2)^\perp]^0 = \mathit{Id}$ (l'opérateur identité).

La base $\mathcal{B}_{Sym}$ optimale est celle qui minimise l'écart entre $\mathbb{C}$ et $\mathcal{F}(\mathbb{C},\mathcal{B})$. Puisque $\mathcal{B}$ est définie par rapport à $\mathcal{B}_O$ à l'aide d'angles d'Euler, la minimisation de l'écart peut se faire par exemple par la méthode du simplex. On obtient ainsi la meilleure approximation $\mathbb{C}_{Sym}$ de $\mathbb{C}$ parmi les tenseurs invariants par $Sym$ et la base associée $\mathcal{B}_{Sym}$. Les groupes de symétrie forment un ensemble ordonné (Cowin, 1987 ; Yong-Zongh, 1991) et peuvent être rangés par niveaux dans l'arbre associé à l'ordre. On peut donc pour chaque niveau déterminer le meilleur groupe de symétrie $Sym$ pour $\mathbb{C}$. Mais le choix du niveau reste arbitraire en fonction des imprécisions sur les mesures et/ou des applications ultérieurement envisagées.

$$\mathbb{C} \begin{vmatrix} 4.49 & 0.45 & 0.63 & 0 & 0 & 0 \\ 0.45 & 13.3 & 0.11 & 0 & 0 & 0 \\ 0.63 & 0.11 & 2.77 & 0 & 0 & 0 \\ 0 & 0 & 0 & 0.52 & 0 & 0 \\ 0 & 0 & 0 & 0 & 0.68 & 0 \\ 0 & 0 & 0 & 0 & 0 & 0.63 \end{vmatrix}$$

Tableau 2. Tenseur $\mathbb{C}_{Sym}$ orthotrope du chêne écrit dans ses axes de symétrie (GPa).
*Table 2. Orthotropic stiffness tensor $\mathbb{C}_{Sym}$ written in it's symmetry axes (GPa).*

Le tenseur orthotrope $\mathbb{C}_{Sym}$ obtenu pour le chêne à partir du tenseur "brut" $\mathbb{C}$ est présenté dans le tableau 2 ; il est cohérent avec les mesures classiques (Guitard, 1988). L'ensemble des écarts relatifs pour chaque type de symétrie est présenté dans le tableau 3. Les exposants indiqués correspondent à la relation d'ordre décrite par Cowin et Mehrabadi. Pour le deuxième niveau de symétrie, l'orthotropie apparaît le meilleur choix pour les symétries du deuxième niveau et l'isotropie transverse pour celles du troisième niveau. Ce résultat est bien sûr attendu pour le matériau bois.

---

[1] $\mathcal{R}[\mathbf{z},2\pi/p](\mathbf{v})$ représente la rotation de $\mathbf{v}$ autour de $\mathbf{z}$ d'un angle $2\pi/p$.

| | Type de symétrie | écart $\mathbb{C} - \mathbb{C}_{Sym}$, (%) | | |
|---|---|---|---|---|
| | Triclinique | 0 | | |
| | Monoclinique | 8,6 | | |
| Orthotrope | 13,2 | Trigonale | 15,5 | |
| Tétragonal | 16 | | | |
| Cubique | 56,3 | Isotrope transverse | 16,6 | |
| | Isotrope | ? | | |

Tableau 3. Ecarts relatifs entre $\mathbb{C}$ et $\mathbb{C}_{Sym}$
*Table 3. Relative discrepancy between $\mathbb{C}$ and $\mathbb{C}_{Sym}$*

**Remarque** : La valeur de p conditionne le niveau de symétrie : monoclinique (p=1 ou 2) ; trigonal (p=3 ou 6) ; orthotrope (p=4) ; isotrope transverse/hexagonal (p=5,7 ou de 9 à l'infini) ; tétragonal (p=8). La symétrie tétragonale peut aussi être obtenue à partir de la symétrie orthotrope en "moyennant" le tenseur orthotrope avec sa rotation de π/2 autour de **z**. De même la symétrie cubique est obtenue à partir de la symétrie orthotrope en "moyennant" le tenseur orthotrope avec les deux rotations qui permutent ses axes orthogonaux.

4. MESURE ULTRASONORE DU TENSEUR DE RIGIDITE

4.1 Deux transducteurs ultrasonores identiques sont disposés de part et d'autre de l'éprouvette polyhédrale à faces parallèles (dont la symétrie matérielle est inconnue). Ils génèrent ([2]) des ondes longitudinales ou transversales. Nous pouvons, pour ces dernières, faire varier la direction θ de polarisation (identique à l'émission et à la réception) en orientant les bras couplés qui portent les transducteurs (fig. 2). Pour obtenir le temps de vol, et donc la vitesse $V_i$ des ondes, chaque signal est comparé par inter corrélation à un signal de référence ([3]).

---

[2] La carte contenue dans un micro-ordinateur P.C. permet d'envoyer un signal de durée très brève dans l'émetteur. Le signal ultrasonore émis a une fréquence centrale de l'ordre du mégahertz.

[3] A partir d'une cale à deux épaisseurs permettant d'établir un zéro temporel.

6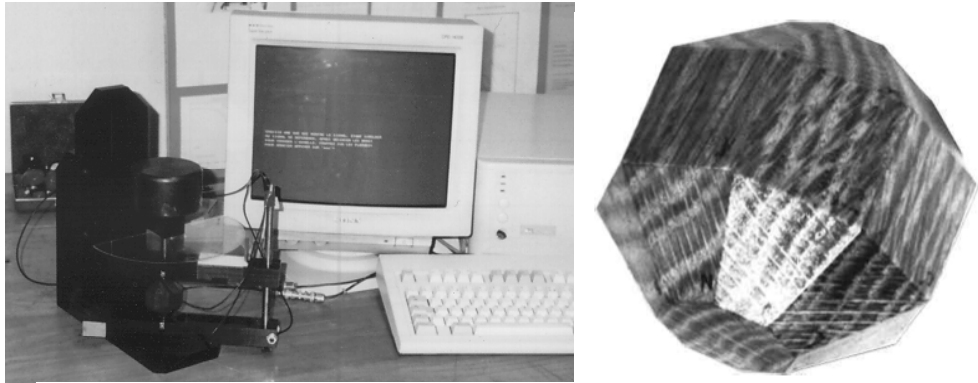

| A | B | C | I | J | K | L | M | N | α | β | γ | δ |
|---|---|---|---|---|---|---|---|---|---|---|---|---|
| 1.0000 | 0 | 0 | 0 | 0.7071 | 0.7071 | 0 | 0.7071 | -0.7071 | 0.5000 | -0.5000 | -0.5000 | 0.5000 |
| 0 | 1.0000 | 0 | 0.7071 | 0 | 0.7071 | 0.7071 | 0 | 0.7071 | 0.7071 | 0.7071 | 0.7071 | 0.7071 |
| 0 | 0 | 1.0000 | 0.7071 | 0.7071 | 0 | -0.7071 | 0.7071 | 0 | 0.5000 | 0.5000 | -0.5000 | -0.5000 |

Fig. 2 - Schéma et repère du montage. La facette X' est opposée à X.
*Fig. 2 - Ultrasonic apparatus and it's coordinates. X' is opposed to X.*

L'émetteur émet une vibration de direction **m** et le récepteur reçoit l'onde propagée dans la direction **n**=**x**$^1$. Les directions de vibration **u**$^i$ sont repérées par les angles de polarisation $\theta_i$ et les angles de déflexion $\varphi_i$ (fig. 3).

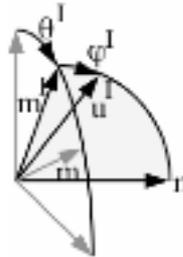

Fig. 3. - Directions de vibration et de mesure.
*Fig. 3 - Vibration directions and measurement directions.*

Pour les ondes quasi-transverses, $\theta_i$ est donné par la recherche de l'intensité maximale. Expérimentalement, seules les trois célérités $V_1$, $V_2$, $V_3$ et les angles $\theta_2$ et $\theta_3$ des polarisations quasi-transversales sont accessibles : l'ensemble est nommé une *mesure complète* (tableau 4).

| Face mesurée | A | B | C | I | J | K | L | M | N | α | β | γ | δ |
|---|---|---|---|---|---|---|---|---|---|---|---|---|---|
| Vitesse Q.L. | 953 | 4387 | 1740 | 2431 | 1671 | 2923 | 2139 | 2202 | 2314 | 2553 | 2673 | 1915 | 2561 |
| Vitesse Q.T. 1 | 1312 | 1403 | 1216 | 1343 | 1172 | 1430 | 1297 | 1388 | 1376 | 1169 | 1505 | 1252 | 1478 |
| Angle 1 | 1 | 62 | 102 | 68 | 4 | 27 | 54 | 60 | 51 | 50 | 64 | 49 | 52 |
| Vitesse Q.T. 2 | 853 | 1151 | 878 | 1064 | 810 | 1011 | 992 | 821 | 978 | 1120 | 951 | - | 955 |
| Angle 2 | 83 | 153 | 10 | 157 | 100 | 154 | 150 | 155 | 150 | 141 | 141 | - | 127 |

Tableau 4. Mesures expérimentales sur le chêne. Vitesses en m/s et angles en degrés.
*Table 4. Experimental measurement on oak. Speeds in m/s and angles in degrees.*

4.2 Le tenseur de Christoffel **Γ** symétrique (Love, 1927), s'écrit en fonction de $\mathbb{C}$ :

(2) $\quad\quad\quad\quad\quad\quad\quad\quad\quad \mathbf{\Gamma} = \mathbf{n}.\mathbb{C}.\mathbf{n}$

$\boldsymbol{\Gamma}$ possède trois vecteurs propres orthogonaux $\mathbf{u}^i$, les directions de vibration des particules, et trois valeurs propres $\rho(V_i)^2$ réelles et positives. Nous pouvons donc l'écrire aussi sous la forme suivante :

$$(3) \qquad \boldsymbol{\Gamma} = \sum_{i=1}^{3} \rho(V_i)^2 \, \mathbf{u}^i \otimes \mathbf{u}^i$$

La première écriture de $\boldsymbol{\Gamma}$ (*équ.* 2) fait intervenir le tenseur $\mathbb{C}$ recherché et la direction de propagation $\mathbf{n}$ connue, tandis que la seconde (*équ.* 3) contient les données expérimentalement mesurées (soulignées) $\underline{V}_i$ et $\underline{\theta}_i$ et les $\varphi_i$ inconnus. Soit $\mathbf{K}$ le tenseur d'écart :

$$(4) \qquad \mathbf{K} = \mathbf{n}.\mathbb{C}.\mathbf{n} - \sum_{i=1}^{3} \rho(\underline{V}_i)^2 \, \underline{\mathbf{u}}^i \otimes \underline{\mathbf{u}}^i$$

Où la direction de vibration quasi-longitudinale $\underline{\mathbf{u}}^1$, non mesurable, est définie comme étant orthogonale aux autres directions de vibrations $\underline{\mathbf{u}}^2$ et $\underline{\mathbf{u}}^3$. Dans le but d'utiliser toutes les données disponibles nous définissons pour chaque mesure la fonction d'écart j suivante :

$$(5) \qquad j = \frac{1}{2} \, \mathbf{K} : \mathbf{K}$$

En minimisant la somme J des E fonctions j (E étant le nombre de mesures complètes), on obtient $\mathbb{C}$ pour $E \geq 6$. La structure de j permet de séparer les variables : les 2E angles de déflexions $\varphi_i$ sont tout d'abord initialisés à zéro (solution pour un tenseur $\mathbb{C}$ isotrope), ensuite la minimisation est faite itérativement par rapport à $\mathbb{C}$ ([4]) puis aux $\varphi_i$. Les 2E angles de déflexion $\varphi_2$ et $\varphi_3$ interviennent de façon indépendante dans chaque terme j qui sera donc minimisé à tour de rôle par un algorithme B.F.G.S.

**Remarque 1.** Van Bursirk (1986) utilise l'expression de $\mathbf{K}$ dans la base ortho. $\mathbf{u}^i$, qui demeure toutefois difficilement mesurable expérimentalement. Contracter $\mathbf{K}$ avec $\underline{\mathbf{u}}^i \otimes \underline{\mathbf{u}}^i$ donne trois équations scalaires qu'Arts (1993) utilise pour déterminer $\mathbb{C}$ ; les $\underline{\mathbf{u}}^i$ inconnus sont initialisés comme $(\mathbf{n},\underline{\mathbf{m}}^2,\underline{\mathbf{m}}^3)$ puis calculés (itérativement) en diagonalisant le tenseur $\boldsymbol{\Gamma}$ issu du $\mathbb{C}$ obtenu ($\underline{\mathbf{m}}^i$ n'est plus en général la projection de $\underline{\mathbf{u}}^i$). La trace de $\boldsymbol{\Gamma}$ et un terme d'écart entre les deux composantes $\Gamma_{11}$ dans la base $\underline{\mathbf{u}}^i$ et une base complétant $\mathbf{n}$ permettent à Klima (1973) de trouver $\mathbb{C}$ à partir de 21 mesures des vitesses seules.

**Remarque 2.** La mesure est *incomplète* lorsqu'un des signaux n'est pas mesurable expérimentalement (tableau 4). Il peut être en effet absorbé par le matériau ou confondu avec des réflexions parasites. Nous définissons alors les composantes $L_{ij}$ obtenue par la projection de $\mathbf{K}$ sur l'espace généré par les directions "connues" $\underline{\mathbf{u}}^i \otimes \underline{\mathbf{u}}^j$ correspondant aux mesures $\underline{\mathbf{m}}^i$ et $\underline{\mathbf{m}}^j$ :

$$(6) \qquad L_{ij} = (\underline{\mathbf{u}}^i \otimes \mathbf{n}) : \mathbb{C} : (\mathbf{n} \otimes \underline{\mathbf{u}}^j) - \delta_{ij} \, \rho \underline{V}_i^2$$

---

[4] Par annulation du gradient ; le calcul est linéaire. Les grandes et petites symétries ($C_{ijkl}=C_{klij}=C_{ijlk}$) sont imposées par 60 multiplicateurs de Lagrange.

Si la mesure quasi-transversale 3 est manquante nous pouvons éliminer l'angle de polarisation $\underline{\theta}_3$ inconnu (deux solutions, on retient la meilleure) en imposant l'orthogonalité du trièdre $\underline{u}^i$. On définit la fonction h suivante :

$$(7) \qquad h = \frac{1}{2} \sum_{i,j} L_{ij}^2$$

Cette définition correspond à h = 1/2 (**L** : **L**) si les $\underline{u}^i$ forment une base orthonormée. La minimisation se fait de même que précédemment. La condition de résolution devient 4E+F≥21 avec F le nombre de mesures incomplètes. Le tenseur ℂ obtenu pour l'éprouvette en chêne est donné dans le tableau 1.

3. CONCLUSIONS ET PERSPECTIVES - Nous proposons une stratégie pour effectuer un choix rationnel du type de symétrie (et la base naturelle correspondante) d'un matériau élastique linéaire dont on connaît seulement le tenseur complet ℂ dans la base de l'éprouvette. D'autres indicateurs pourraient être issus d'une décomposition irréductible de ℂ (Bœhler, 1994 ; Surrel, 1993) ou de la décomposition de Kelvin (Rychlewski, 1984). Pour tout essai ultrasonore où l'on connaît la direction de propagation **n** la méthode d'identification de ℂ peut être utilisée. Si la symétrie de ℂ est connue *a priori*, d'autres méthodes ultrasonores sont à préférer.

6. BIBLIOGRAPHIE